\documentclass[preprint2]{aastex}

\def\eg{{\it e.g.,} }
\def\etal{{\it et al.}}
\def\ie{{\it i.e.,} }
\def\lsim{\lower.5ex\hbox{$\; \buildrel < \over \sim \;$}}
\def\gsim{\lower.5ex\hbox{$\; \buildrel > \over \sim \;$}}
\def\simeq{\lower.3ex\hbox{$\; \buildrel \sim \over - \;$}}

\slugcomment{draft of \today}

\shorttitle{Quasi-Spherical, Time-Dependent Viscous Accretion Flow}
\shortauthors{Lee et al.}

\begin{document}

\title{QUASI-SPHERICAL, TIME-DEPENDENT VISCOUS ACCRETION
FLOW: ONE-DIMENSIONAL RESULTS}

\author{Seong-Jae Lee$^{1}$, Dongsu Ryu$^2$, and Indranil Chattopadhyay$^3$}
\affil{$^1$ School of Science Education, Chungbuk National
University, Chungbuk 361-763, Korea seong@chungbuk.ac.kr\\
 $^2$Department of Astronomy and Space Science,
Chungnam National University, Korea\\
$^3$ ARIES, Manora Peak, Nainital-263129, Uttarakhand, India \\}

\begin{abstract}
We investigated the instability of advective accretion flow as a consequence of angular momentum transfer in one-dimensional, quasi-spherical transonic accretion flow around a non-rotating black hole. The code is designed to include the effects of viscosity; the hydrodynamics component preserves angular momentum strictly with Lagrangian and remap method in absence of viscosity, while the viscosity component updates viscous angular momentum transfer through the implicit method. We performed two tests to demonstrate the suitability of the code for accretion study. First, we simulated the inviscid, low angular momentum, transonic accretion flow with shocks around a black hole, and then the subsonic, self-similar ADAF solution around a Newtonian object. Both simulations fitted the corresponding analytical curves extremely well. We then simulated a rotating, viscous, transonic fluid with shocks. We showed that for low viscosity parameter, stable shocks at larger distance are possible. For higher viscosity parameter, more efficient angular momentum transfer in the post-shock disk makes the shock structure oscillatory. Moreover, as the shock drifts to larger distances, a secondary inner shock develops. We showed that the inner shock is the direct consequence of expansion of the outer shock, as well as creation of regions with $\partial l / \partial r < 0$ due to more efficient angular momentum transfer near the inner sonic point. We showed that all disk parameters, including emissivity, oscillate with the same period as that of the shock oscillation. Our simulation may have implication for low frequency QPOs, e.g., GRO J1655-40 and XTE J1550-564.
\end{abstract}




\section{INTRODUCTION}

Investigation of the flow behavior of the accreting matter in the vicinity of a black hole is important
since the spectrum and the intensity of the emitted radiation depend on the flow structure. The event horizon
presents the unique inner boundary condition in which the in-falling matter crosses the horizon with the speed of light ($c$).
Therefore black hole accretion has to be transonic, as a result of which existence of one sonic point or critical point is assured for black hole accretion. General relativity also ensures that
matter must posses sub-Keplerian angular momentum closer to the horizon. Although within the marginally stable circular orbit ($r_{ms}$), the angular momentum at $r{\lsim}r_{ms}$ is definitely sub-Keplerian (and the value $l {\sim} l_{ms}=l|_{r=r_{ms}}$), but at larger radius the  angular momentum should be generally large.
Therefore, a general accretion disk should have viscosity to remove the angular momentum outwards.
The first serious model of viscous accretion disk was presented by \citet{ss73}, in which the angular momentum distribution was Keplerian, the accretion disk was geometrically thin and optically thick. In Shakura-Sunyaev disk the pressure and advection terms were not properly considered and no attempt was made to
satisfy the inner boundary condition around a black hole apart from the adhoc termination of the disk at
$r \leq r_{ms}$. Along with this theoretical short coming,
Shakura-Sunyaev disk also failed to explain power-law high energy part of a black hole candidate spectrum. Therefore, search for another component of an accretion disk which may explain the origin of the high energy radiations from black hole candidates, were undertaken by various groups.
One such model which got a wide attention was ADAF [\eg \citet{i77}, \citet{ny94} hereafter NY94]. This model was first constructed around a Newtonian gravitational potential, where the viscously dissipated energy is advected along with the mass, momentum and the entropy of the flow. The original ADAF solution was self-similar and wholly subsonic, and was found to be thermally and dynamically stable. Howover, the low viscosity ADAF showed convective instability \citep{ia99}, that has no dynamical effect if the angular momentum is transported outward but it is dynamically important in case the opposite is true. The global solution of ADAF showed that the flow actually becomes transonic at around few Schwarzschild radii ($r_g$), and the self-similarity may be maintained far away from the sonic point \citep{cal97}.

Simultaneous to these developments, there were some interesting research going on sub-Keplerian flows around black holes.
It has been shown that sub-Keplerian flow does posses multiple sonic point in a significant range of the energy-angular momentum parameter space \citep{lt80}. One of the consequences of existence of multiple sonic points, is that the flow accreting through the
outer sonic point can be slowed down by the centrifugal barrier. This slowed down matter acts as barrier to the faster fluid
following it. If the strength of the barrier is strong enough then accretion shocks may form \citep{c89}. General global solutions in the advective domain incorporating viscosity and thermal effects were obtained by many independent researchers \citep{c90,c96,lgy99,lmc98,gl04}.
Furthermore, it has also been shown that the global ADAF solution is a subset of the general advective solutions \citep{lgy99}. Whether a flow will follow an ADAF solution or some kind of hybrid solution
with or without shock will depend on the outer boundary condition and the physical processes dominant in the disk.

Although steady-state solutions are possible in a certain range of parameter space \citep{c89,cd04,mlc94,mrc96a}, but advective solutions with discontinuities such as shocks
are generally prone to various kind of instabilities. Since, various flow variables across the shock surface
jumps abruptly, this results in a markedly different cooling, heating and other dissipation rates across the shock. This may render the shock unstable. For example, in presence of bremsstrahlung cooling, resonance between cooling timescales and in fall timescales in the post shock part of the disk gives rise to oscillating shocks \citep{msc96b}. \citet{lmc98} showed that beyond a critical viscosity post-shock disk may oscillate. The interaction of the outflow and the inflow
may also cause the bending instability in the disk \citep{makbc01}. \citet{mtk99} showed that in presence of non-axisymmetric azimuthal perturbations the shock initially becomes unstable but stabilizes within a finite radial extent into an asymmetric closed pattern.
Moreover, the post-shock region may be associated with the elusive Compton cloud that produces the hard photons \citep{ct95,cm06,mc08} and may also be the base of the jet \citep{dc99,dcc01,cd07,dc08,bdl08,dbl09}. Therefore instabilities
of the post-shock region may manifest itself as the variabilities observed in the emitted hard photons seen in microquasars and active-galactic nuclei \citep{msc96b}. To add a new twist, \citet{ft04} conjectured the presence of multiple shocks and \citet{gcsr10} actually reported the presence of two oscillating shocks giving rise to two quasi-periodic oscillations.

In this paper, we concentrate on the study of instabilities of rotating fluid around black holes, generated by the angular-momentum transport by viscosity. Since the temperature, density etc are higher and the velocity is lower in the post-shock region compared to the pre-shock region, the angular momentum transport rate should be different
in the two regions of the disk. In other words, in this paper we simulate transonic, viscous, rotating
fluid around black holes. We employ a new code to study the effect of angular momentum transport in the accretion disk. Unlike other purely Eulerian codes, this new code is especially developed to strictly conserve angular momentum in absence of viscosity. In \S 2, governing equations and assumptions are presented. In \S 3, the code which was built to calculate the evolution of angular momentum as accurately as possible is described, along with tests for a rotating transonic flow and a viscous flow. In \S 4, the structure and the instability shown in simulations are presented, along with descriptions on the nature of the instability. A summary and discussion is presented in \S 5.

\section{BASIC EQUATIONS}

The one-dimensional time-dependent equations for quasi-spherical
accretion of viscous flows are given by
\begin{equation}
{\partial \rho \over \partial t} + {1 \over r^2}
{\partial \over \partial r} (r^{2}\rho v_r) = 0,
\end{equation}
\begin{equation}
{\partial v_r \over \partial t} + v_r{\partial v_r \over \partial r} +
{1 \over \rho}{\partial p \over \partial r} = {l^2 \over r^3} -
{\partial \Phi_i \over \partial r},
\end{equation}
\begin{equation}
{\partial l \over \partial t} + v_r {\partial l \over \partial r} =
{1 \over r^2 \rho} {\partial \over \partial r} \left[\mu r^4 {\partial
\over \partial r} \left({l \over r^2}\right)\right],
\end{equation}
\begin{equation}
{\partial e \over \partial t} + v_r {\partial e \over \partial r} +
{p \over r^2 \rho}{\partial \over \partial r}(r^2 v_r) = f {\mu r^{2}
\over \rho} \left[{\partial \over \partial r} \left({l \over r^2}\right)\right]^2,
\end{equation}
where $\rho$, $v_r$, $l$, $\Phi_i$ and $e$ are the gas density, radial velocity,
specific angular momentum, gravitational potential and specific internal energy, respectively.
The angular velocity is defined as $\Omega = l/r^2 $. The suffix $i$ in equation (2) denotes ${\rm N}$
or ${\rm PN}$, corresponding to Newtonian or pseudo-Newtonian gravity \citep{pw80}, respectively,
and are given by
\begin{equation}
\Phi_{\rm N} = -{GM_{BH} \over r}
\end{equation}
and
\begin{equation}
\Phi_{\rm PN} = -{GM_{BH} \over r - r_g}
\end{equation}
where $M_{BH}$ is the black hole mass and the Schwarzschild radius is $r_g=2GM_{BH}/c^2$.
The pseudo-Newtonian potential is widely used to mimic the Schwarzschild geometry.
For the gas pressure the equation of state for ideal gas is assumed,
\ie\
\begin{equation}
p = (\gamma-1) \rho e,
\end{equation}
where $\gamma$ is the ratio of specific heats.
For viscosity, the $\alpha$ prescription
(Shakura \& Sunyaev 1973) can be assumed, \ie\ the dynamical
viscosity coefficient is described by
\begin{equation}
\mu = \alpha \rho {c_s^2 \over \Omega_K},
\end{equation}
where
\begin{equation}
c^2_s=\frac{\gamma p}{\rho}
\end{equation}
is the square of the adiabatic sound speed, and
\begin{equation}
\Omega_K = {l_K \over r^2}= \left[\frac{1}{r}\frac{d\Phi_i}{dr}\right]^{1/2}     
\end{equation}
is the Keplerian angular velocity, and the viscosity parameter $\alpha$ is a
constant which is generally less than 1. It is to be noted that the actual expression of
$\Omega_K$ depends on the gravitational potential used. Finally following NY94, the
parameter $f$ measures the fraction of the viscously generated
energy that is stored as entropy and advected along with flows.
The value $f=1$ corresponds to the limit of full advection and has been used in this paper.

In the following, we use $c$ and
$r_g$ as the units of velocity and length,
respectively, unless otherwise stated. In the geometrical units, the unit of time is $\tau_g=r_g/c$.

\section{CODE AND TESTS}

One of the most demanding tasks in carrying out numerical
simulations of equations (1) -- (4) is to calculate the evolution of the
angular momentum as accurately as possible. Capturing shocks
sharply should also be important in resolving structures with clarity, if
shocks are involved. It has been known that the latter can be
achieved by using codes based on modern, upwind finite-difference
schemes on an Eulerian grid. However, without viscosity in such Eulerian codes,
it is normally the azimuthal momentum ($\rho
v_\phi$) and not the angular momentum that is treated as a conserved
quantity. On the other hand, codes based on the Lagrangian concept,
such as the SPH code, can be designed to preserve the angular
momentum strictly. Although it has been successfully applied to many
studies of accretion flows however, the SPH code is known to be
unduly dissipative 
\citep[see, \eg][for discussions]{mrc96a}.

Here, we describe an Eulerian code that was built to accurately calculate
the evolution of the angular momentum including its
transport due to viscosity, and at the same time to capture
discontinuities (shocks and contact discontinuities) sharply with
minimum numerical dissipation. The code is composed of two parts;
hydrodynamic and viscosity part. The hydrodynamic part is based
on the Lagrangian TVD plus remap approach. The Lagrangian/remap
approach is not new in numerical hydrodynamics and was employed
previously \citep{cw84}. But here we show
that in this approach the equation for angular momentum
conservation can be directly solved, so the hydrodynamics part can
be designed to preserve the angular momentum strictly in the absence
of viscosity. At the same time, the TVD scheme \citep{h83,rokc93} guarantees sharp reproductions of discontinuities and
minimum numerical dissipation. In the viscosity part, the viscous
angular momentum transfer is updated through an implicit method, assuring
it is free from numerical instabilities related to it. But the
viscous heating is updated with a second order explicit method,
since it is less subject to numerical instabilities.

\subsection{Hydrodynamic Part}

The hydrodynamic part consists of the Lagrangian step and the remap step.
First in the Lagrangian step, the equations for Lagrangian
hydrodynamics are solved.
On the Lagrangian grid defined with mass coordinate, equations (1) -- (4),
except for the centrifugal force, gravity, and viscosity terms which are
treated separately [see below], can be written in a conservative form as
\begin{equation}
{d\tau \over dt} - {\partial(r^2 v_r) \over \partial m} = 0,
\end{equation}
\begin{equation}
{dv_r \over dt} + r^2{\partial p \over \partial m} = 0,
\end{equation}
\begin{equation}
{dl \over dt} = 0,
\end{equation}
\begin{equation}
{dE \over dt} + {\partial(r^2 v_r p) \over \partial m} = 0,
\end{equation}
where ${\tau}$ and $E$ are the specific volume and the specific total
energy, respectively, that are related to the quantities used in equations
(1) -- (4) as
\begin{equation}
\tau = {1 \over \rho}, \qquad E = e + {v^2_r \over 2}.
\end{equation}
The mass coordinate is related to the spatial coordinate via
\begin{equation}
dm = \rho(r) r^2 dr,
\end{equation}
and its position can be followed with
\begin{equation}
{dr \over dt} = v_r(m,t).
\end{equation}
Equations (11), (12), (14) form a hyperbolic system of conservation
equations, and upwind schemes can be applied to build codes that
advance the Lagrangian step using the Harten's TVD scheme, which is
an explicit, second-order, finite difference scheme to
solve a hyperbolic system of conservation equations \citep{h83,rokc93}. We note that the angular momentum in Eq. (13) is
preserved, so it need not be updated in the Lagrangian step.

In the remap step, the quantities evolved in the Lagrangian grid
are redistributed to the Eulerian grid, to preserve the spatially fixed
grid structure.
Before the Lagrangian step, the Lagrangian and Eulerian grid zones coincide.
But after the step, the Lagrangian grid zone moves to the updated
position
\begin{equation}
r^{\rm new} = r^{\rm old} + {\bar v} \Delta t,
\end{equation}
where $\bar v$ is the time-averaged velocity, and so it does not
coincide with the Eulerian grid zone any more. Not only are the
quantities conserved in the Eulerian grid, the density, radial
momentum, and total energy, but also the angular momentum are
remapped. For the remap, we employ the third order accurate scheme
used in the PPM code (see Colella \& Woodward 1984, for details).

With the Lagrangian and remap steps, equations (1) -- (4) are updated in
the Eulerian grid, except for the centrifugal force, gravity, and viscosity
terms on the right hand side.
The centrifugal force and gravity terms are calculated separately after
the Lagrangian and remap steps such that
\begin{equation}
v_i^{\rm hydro} = v_i^{\rm lag+remap} + \Delta t \left( {l_i^{\rm remap}
\over r_i^3} - {d\Phi \over dr}\Bigg|_i \right).
\end{equation}
Then the viscosity terms are calculated, as discussed in the next
subsection.

Non-uniform Eulerian grids can be employed in the code. For the
problem in this paper we use a grid, where the size of cells
increases exponentially as
\begin{equation}
\Delta r_i = \Delta r_1 \times \delta^{i-1},
\end{equation}
to achieve higher resolution at the origin.
Here $\Delta r_1$ the size of the first grid cell and $\delta$ is
the increment factor.

\subsection{Viscosity Part}

Viscosity has two effects on accretion flows.
First, it transfers the angular momentum outwards, allowing the matter
to accrete inwards.
At the same time, it acts as friction, which results in viscous heating.

Since the term for the angular momentum transfer in Eq. (3) is linear
in $l$, it can be solved implicitly.
Substituting of $(l^{\rm new} + l^{\rm remap})/2$ for $l$, Eq. (3) without the
advection term becomes
\begin{equation}
a_i l_{i-1}^{\rm new} + b_i l_{i}^{\rm new} +c_i l_{i+1}^{\rm new} =
-a_i l_{i-1}^{\rm remap} - (b_i -2)l_{i}^{\rm remap} -c_i l_{i+1}^{\rm remap},
\end{equation}
forming a tridiagonal matrix. Here $a_i$, $b_i$, and $c_i$ are given
with $\rho$, $\mu$, and $r$ as well as $\Delta r$ and $\Delta t$.
The tridiagonal matrix can be solved easily for $l^{\rm new}$ with
an appropriate boundary condition \citep{ptvf92}. The
term for the viscous heating in Eq. (4) is also linear in $e$ (note
that $\mu \propto e$), so it alone can be solved implicitly too.
However, combining the two linear equations for $l$ and $e$ becomes
complicated. Through numerical experiments, we found that the
explicit treatment for the viscous heating does not cause any
numerical problem. So instead of implementing a complicated scheme
to solve simultaneously $l$ and $e$ implicitly, we solve the angular
momentum transfer implicitly, while solving the viscous heating
explicitly.

\subsection{Tests}

Two tests are presented to demonstrate that the code can handle transonic
flow as well as viscous flow which are involved in our problem.
The capability of the code to capture shocks sharply and resolve
structures clearly is tested with a transonic accretion flow. In these tests we reproduce well-known results.

The evolution of an inviscid flow, which enters the outer boundary with
a small amount of angular momentum and approaches a black hole
described by Paczy\'nski \& Wiita potential \citep{pw80}, is calculated in cylindrical geometry. We note that previous subsections describe the
code only in spherical geometry, but the code is actually
written in arbitrary geometries. For this test, the version in the cylindrical geometry is used. Without viscosity, the angular
momentum is preserved. A shock can form, if the rotating flow through the outer sonic point
approaches the centrifugal barrier and decelerates discontinuously, due to the twin effect of centrifugal force
and pressure.
In the test, $l=1.8cr_g$, a value slightly below the
marginally stable value [$l_{ms}=(3/2)^{3/2}cr_g$], is used. The values of the flow
quantities in geometrical units are $(\rho, p, v_r) = (0.71809, 0.007604, -0.083566)$ at
the injection radius $r_{\rm inj}=50r_g$. The sound speed at $r_{\rm inj}$ is $c_s=0.1188c$, hence the fluid is subsonically injected. The fluid becomes supersonic after crossing the outer sonic point
at $r_{co}=27.9r_g$, becomes subsonic at the shock $r_{sh}=7.89r_g$ and enters the blackhole supersonically
after crossing the inner sonic point at $r_{ci}=2.563r_g$. For the ratio of specific
heats, $\gamma = 4/3$ is used. This is the case of a stable standing
shock considered in \citet{mrc96a}. In Figure 1, the
numerical solutions (open circles) of density $\rho$ (top) and radial Mach number $M_r=v_r/c_s$ (bottom) using 2048 uniform grid cells is compared with
the analytical solution (solid line). The flow in smooth
regions coincides with the analytical solution very well and the
shock position matches the analytical value very well. The agreement of analytical result
with the current code is better than those with the purely Eulerian
TVD code and the SPH code presented in \citet{mrc96a}.

Next, the performance of the code for a subsonic viscous flow is tested with a self-similar ADAF solution. Matter is steadily injected with Keplerian angular velocity
into the computational domain at $r_{\rm inj}$, and the simulation
lasts until the steady state is reached. Figure 2 presents the flow
quantities after the steady state is reached and compares them with
the analytic solution. The Newtonian potential is used, and the ADAF
is described with a self-similar solution (NY94). The values of
physical parameters used in this test are $\gamma = 4/3$, and $\alpha =
0.3$. The simulation was performed on an exponentially
increasing grid of 780 cells with $\Delta r_1 = 0.4972 r_s$ and
$\delta = 1.01$. Here $r_s$ is the sink size. The injection position
is $r_{\rm inj} \sim 3.6915 \times 10^4 r_s$. The quantities are
drawn in units of the Keplerian velocity and the Keplerian angular
momentum at the sink, $v_K(r_s)$ and $l_K(r_s)$, and the density is
in an arbitrary unit. The figure shows that the analytic solution is
reproduced very closely in the region between $r \sim 10 r_s$ and $r
\sim 10^4 r_s$ in a box of size $1.1611 \times 10^5 r_s$. The error
in the specific angular momentum is less than a few \% at most.

\section{RESULTS OF SIMULATIONS}

In previous numerical works, oscillation phenomena in accretion flows around black holes related to the QPO were reported. The study of
inviscid supersonic accretion flows around a Newtonian central object, showed that the accretion disk with
shock structure to be dynamically unstable \citep{rbol95}.
Global transonic accretion flows around black holes have been known to exhibit stationary shocks for inviscid
\citep{c89,mrc96a,dcc01} as well as dissipative flows \citep{c96,lgy99,lmc98,bdl08}.
However, since the post-shock flow is hotter, denser and slower, the dissipation rate in
the post-shock flow is shorter than that of the pre-shock flow, which may make the post-shock flow
unstable.
Indeed it has been shown that the energy-angular momentum parameter space for standing shock decreases with the increase of
viscosity parameter \citep{cd04,gl04,dbl09}. \citet{lmc98} simulated viscous transonic flow and showed steady shocks exist for low viscosity, while for higher viscosity shock becomes unstable.
However, \citet{lmc98} restricted their investigations for a hot flow ($T_{inj}\sim 10^{11}K$
at the injection radius), and very low viscosity
parameter ($\alpha \lsim$ few$\times10^{-3}$). Since the pre-shock flow was chosen to be hot (post-shock disk was obviously even hotter), the angular
momentum removal was very efficient in both the pre-shock as well as post-shock disk, even when the viscosity parameter was low. The length scale of the computation box was only about few tens of $r_g$.
In this paper we simulate viscous transonic flow which is cold to begin with, and investigate the instability arising from
reasonably higher viscosity of the flow. The reason to choose cold flow at the injection is to
have a very different angular momentum transport rate in the post-shock and pre-shock disk,
and thereby to maximize the effect of shock instability.
Moreover, we keep the length scale of computation box fairly large
so as to study large amplitude and low frequency shock instability.

\subsection{Shock Formation in Inviscid Rotating Flow}

We start our set of simulations with a low energy, rotating, transonic, inviscid flow around a black hole (described by $\Phi_{PN}$).
The steady-state, inviscid, transonic solution corresponds to a flow characterized by Bernoulli parameter or specific energy [which in our unit system is ${\cal E}=0.5v^2_r+c^2_s/(\gamma -1)+l^2/(2r^2)-1/\{2(r-1)\}$] and specific angular momentum ($l$). Parameters ${\cal E}, ~l$ for the inviscid flow
are $1.25{\times}10^{-6}c^2$ and $1.8cr_g$, respectively. The ratio of specific heat is given by $\gamma=1.4$. The Bondi radius (the length scale within which gravity becomes important) is defined as
$r_B = GM_{BH}/c_{s,\infty}^2$, where $c_{s,\infty}^2={\cal E}(\gamma-1)$ for an inviscid flow.
In this particular case $r_B=10^6r_g$ and $c_{s,\infty} =  7.071~{\times}10^{-4}c$. The analytical, steady-state solution of flows for these parameters gives two physical sonic points, the inner one is at
$r_{ci}= 2.394r_g$ and the outer one at $r_{co}=199991.04r_g \approx 0.2r_B$. The analytical solution also predicts a shock at $r_{sh}=22.2r_g$. It has been shown in connection to Figure 1, that it is possible to simulate transonic flow quite accurately with subsonic injection \ie when  $r_{inj}>r_{co}$.
However in the present scenario, $r_{sh}\ll r_{co}$. Therefore, if $r_{inj}>r_{co}$, then a large amount of computation time will be wasted in simulating uninteresting region
of the disk. Hence, without any loss of generality, we choose the injection parameters from the supersonic portion of the analytical curve in order to reduce computation time.
To further reduce the computation time and  also to
achieve higher resolution close to the center, we use
exponentially-increasing grids, which are 3,553 cells with $\Delta
r_1 = 0.0296$ and $\delta = 1.001$ and the length of the computation box corresponds to $1000 r_g = 0.001 r_B$.
The injection radius is hence $r_{inj}=1000r_g=0.001r_B$, and
the flow radial velocity [$v_r (inj) = 2.970 \times 10^{-2}c$], specific angular momentum [$l_{inj}=1.8cr_g$] and sound
speed [$ c_s(inj) = 4.827 \times 10^{-3}c$] at $r_{inj}$ is taken from the analytical solution.
In Figure 3, we compare the steady-state analytical solution (solid) with simulation result (open circle) when steady state is reached. Various flow variables like $\rho$ (a), $v_r$ (b),
$c_s$ (c) and $p$ (d) are plotted with $log(r)$. Figure 3, shows excellent agreement of the simulation result with the analytical curve and the shock has been captured within 2-3 cells.

\subsection{Shock Oscillation of Viscous Flow}

Time dependent solutions of viscous transonic accretion flow are obtained by starting with the inviscid flow described in
\S 4.1 as the initial condition, and then increasing the viscosity parameter $\alpha$.
The action of the viscosity can be understood from equations (3,4). For accretion (\ie
when $v_r < 0$), if the r.h.s of equation (3) is negative then the
angular momentum will be transported outwards. 
In presence of Shakura-Sunyaev type viscosity, the
angular momentum transport in the post-shock subsonic flow is much more efficient than the pre-shock
supersonic flow. As a result of which, angular momentum starts to pile up in the immediate post-shock fluid, resulting in a jump in the angular momentum distribution across the shock.
Similarly, equation (4) tells us that the viscous heat dissipation in the post shock disk will also be higher compared to the pre-shock disk.

It is well known that a standing shock forms if the total pressure (ram+gas pressure)
is conserved across the shock \citep{c89}.
In Figure 4a, the inviscid solution produces a stationary shock location $r_{sh}=22.2r_g$, as is
shown in Figure 3.
In Figure 4b, the shock location as a function of $t$ is plotted for $\alpha=0.003$. The excess gas
pressure due to viscous heat dissipation, and the increased centrifugal force due to the piled-up
angular momentum in the post-shock disk, pushes the shock front outwards.
For low $\alpha$, shock front moves to a larger location ($r_{sh}\sim 31r_g$ as in
Fig. 4b) where the balance between
total outward push and the total inward pressure from the pre-shock flow is restored.
However, for higher viscosity parameter $\alpha=0.006$ (Fig. 4c),
the enhanced angular momentum transport creates an even more stronger outward push
and the shock front  overshoots a possible equilibrium position and the shock starts to oscillate. Interestingly, when the shock moves to around $\sim 70~r_g$ and beyond, a second shock tends to emerge, which expand and collide with the outer shock. The combined shock then drifts outwards, the inner shock re-emerges and the cycle continues. In the following, let us make a detailed investigation on transonic flow with higher viscosity, and the emergence of two shocks.

In Figure 5a -- 5d, we have plotted radial velocity $v_r$ (dashed-dot) and the sound speed $c_s$ (solid)
at four time steps (a) $t=2.9{\times}10^5 r_g/c$, (b) $t=3.165{\times}10^5r_g/c$, (c) $t=3.4{\times}10^5r_g/c$
and (d) $t=3.615{\times}10^5r_g/c$, where the outer boundary conditions are same as Figure 3, and $\alpha=0.01$.
In Figure 6a -- 6d, the specific angular momentum distribution $l(r)$ is plotted for the same simulation and for the same time steps as in Figure 5a -- 5d. Corresponding panels of Figures 5 and 6 are to be considered in tandem to understand this complicated phenomenon.
The different snap shots in Figures 5a -- 6d, corresponds to (i) the maxima in outer shock (Figs. 5a and 6a),
(ii) the expansion stage of the combined shock just after the minima in the next cycle (Figs. 5b and 6b), (iii) just before the maxima of the outer shock (Figs. 5c and 6c), and (iv) just after the maxima
of the outer shock (Figs. 5d and 6d).
In Figure 5a, there are two shock structures, the inner shock is at $r_{sh}({\rm in})\sim 130r_g$ and the outer
shock $r_{sh}({\rm out})\sim 500r_g$. Corresponding angular momentum distribution in Figure 6a shows that the
$dl/dr>0$ in the range $r<20r_g$ and $r_{sh}({\rm in})<r<r_{sh}({\rm out})$, while $dl/dr<0$ is in the range
$20r_g<r<r_{sh}({\rm in})$. In Figure 5b, the two shocks merge and the combined shock is at $r_{sh}({\rm in})=r_{ sh}({\rm out})=r_{sh}\sim100r_g$. Figure 6b shows that $dl/dr>0$ in a region where $r<20r_g$, and $dl/dr<0$ for $20r_g<r\lsim r_{sh}$, with a smaller hump in angular momentum distribution around $r_{sh}$.
The angular momentum distribution attains a tall peak, and the enhanced centrifugal pressure almost stalls the infall (Fig. 5b) in that region.
However, due to the extra pressure from the piled-up $l$ the combined shock moves outwards, while the contact discontinuity wave resulting from the collision of shocks moves towards the black hole.
As a result the sound speed (\ie temperature) in the immediate post-shock flow drops (Fig. 5c), and the angular momentum distribution becomes $dl/dr>0$ (Fig. 6c). This allows for a more free infall and $v_r$ in the immediate post shock disk increases considerably (Fig. 5c). As the contact discontinuity wave is absorbed by the black hole the angular momentum in the immediate post-shock region gets reduced considerably, $v_r$ becomes supersonic in the region. However, the flow again hit the centrifugal barrier closer to the black hole and the inner shock reemerges (Figs. 5d and 6d). We note that the regions of $dl/dr<0$ are subject to the rotational instability. The non-steady behavior shown here should be partly attributed to the instability.

\subsubsection{On emergence of the inner shock, shock collision and the angular momentum transfer}

In the top panel of Figure 7, the shock oscillation is plotted for $\alpha=0.01$. Therefore, each of the panels of Figures. 5a -- 6d corresponds to the various snap shots of flow variables taken from the top panel of Figure 7 (time sequences a -- d
are marked on the figure).
Similar to Figure 4c, $r_{sh}$ in the top panel of Figure 7 starts to oscillate as the viscosity is turned on.
A transient inner shock, \ie $r_{sh}({\rm in})$ develops when  $r_{sh}({\rm out})\gsim80r_g$.
Initially the dynamics of the two shocks are similar to that of Figure 4c, \ie the inner shock forms
when $r_{sh}({\rm out})$ is at the maxima, and then $r_{sh}({\rm in})$ collides with the contracting outer shock, and the merged shock then expands. However, for $t> 0.2{\times}10^5\tau_g$ the shock dynamics slowly change, both shock expands and then contracts, and the shocks collide while contracting.
The merged shock then reaches a minima and then expands, and this cycle continues.
The query about the formation of the inner shock can be understood as follows:
as the original shock expands to a distance $\gsim 80r_g$, the sound speed in the immediate postshock region and close to the black hole differs by almost an order of magnitude.
Hence the rate of angular momentum transport in a region nearer to the inner sonic point is much higher than the region closer to the shock. Hence, the angular momentum transport rate is not only markedly different
between the post-shock and pre-shock region, but also within the post-shock flow when the shock expands to a very large distance. Hence the angular momentum piles up in between inner sonic point ($r_{ci}$) and the shock (\eg Fig. 6b), which enhances the centrifugal barrier and impedes the accretion.
Continued shock expansion reduces the post shock sound speed (\ie temperature) and creates a mild but positive angular momentum gradient, which increases the infall velocity in the immediate post-shock
flow. This can continue up to the extent that the {\it post shock fluid once again become supersonic
in the immediate post-shock domain, however further down-stream the piled-up angular momentum virtually stops the supersonic inflow causing the formation of the inner shock}. The inner shock again jacks up the temperature, which causes the inner shock to expand too. If the outer shock is contracting
then the two shocks may collide, or, both the shock may expand in phase and collide during the
contraction phase. The combined shock then expands and the whole cycle is repeated. It may be noted that the inner shock emerges half way into each of the cycles and hence it is a persistent feature.

It is interesting to seek the radiative property of such oscillatory dynamics of the disk.
We estimate the bremsstrahlung loss aposteriori from the disk, as a representative of the radiative
loss.
It is well known that the bremsstrahlung emission $\propto \rho^2~c_s$. In Figure 7b,
we plot $E_{Total~Br}/\rho_{inj}^2~c_{s}(inj)$ as a function of time, where
\begin{equation}
 E_{Total~Br} =\int^{r_{inj}}_{r_{sink}}\rho^2~c_s~r^2dr,
\end{equation}
 $\rho_{inj}$ and $c_{s}(inj)$ are the density and sound speed of the flow at the outer boundary or $r_{inj}$.
Hence, the bottom panel of Figure 7 represents total bremsstrahlung emission from the computational box compared to the bremsstrahlung emission at $r_{inj}$. Interestingly, bremsstrahlung emission also has a periodic behavior, whose period is similar to the period of the shock oscillation. However, the shock maxima/minima may or may not coincide with either emission maxima or minima. In this particular case, initially there is no distinct correlation, but as the oscillation approaches quasi saturation, the emission maxima coincides with the combined shock minima. And the emission minima coincides with the rising phase of the combined shock and when the inner shock has not been formed.
As the combined shock contracts it pushes the post-shock matter inward just like a `bellow' in a blacksmiths
shop. This jacks up the $\rho$, $c_s$ and $v_r$. The enhanced $\rho$ and $c_s$ contribute to form the
emission maxima. As the combined shock expands, the flow variables like $\rho$, $c_s$ in the immediate post region decreases, and the emission starts to decrease until it reaches the minima. Since the wide difference of $c_s$ also triggers the differential
$l$ transfer in the inner regions and outer regions of the post-shock disk, the angular momentum again starts to pile up and starts the formation of the inner shock described above.
There exists a secondary peak in the bremsstrahlung emission as well which appears to be related to the dynamics of the inner shock.
The time lag between the shock maxima and the emission maxima is $\delta t\sim 2{\times}10^4\tau_g$.
Initially the oscillation period of the shock was $T^{\prime}_{osc}\sim 5\times 10^3 \tau_g$.
The oscillation period gradually increases to a quasi-saturation value of $T_{osc}\sim 8 {\times} 10^4 \tau_g$.
Since,
\begin{equation}
\tau_g = {2GM \over c^3} \sim 10^{-5} {M_{BH} \over M_\odot}\ {\rm
sec},
\end{equation}
therefore,
\begin{equation}
T_{\rm osc} \sim 8 \times 10^4 \tau_{g} \sim 8 \times 10^{-1}
~ {M_{BH} \over M_{\odot}}\ {\rm sec}.
\end{equation}

This would correspond to the frequency $\sim {0.125 Hz}$ for stellar mass black hole \ie $M_{BH}\sim 10 M_{\odot}$.
In case of a super-massive black hole ($M_{BH}\sim 10^8 M_{\odot}$), these time-scales will
correspond to $2.5$ yr variabilities.
However, since there are two shocks, we are interested to see the influence of the dynamics of the two
shocks on emission.
In the left panel of Figure 8, we plot the outer shock (top), inner shock (middle) and the
relative bremsstrahlung emission (bottom) for reference, and in the right panels
we plot the power density spectra for the outer shock (top), inner shock (middle) and the bremsstrahlung
emission (bottom) for a stellar mass black hole ($M_{BH}=10M_{\odot}$). The power density spectrum of the outer shock shows a frequency of $\sim {0.125 Hz}$. The power density spectrum of the inner shock has a prominent peak at the frequency $\sim 0.25 Hz$
and a weaker peak around $\sim 0.125 Hz$, the secondary peak suggests that the oscillation of the outer shock forces a weak periodicity on the inner shock as well. The power density spectrum of the inner shock is a bit noisy, since the time variation of the inner shock, although persistent, is not continuous. Interestingly, Figure 8, shows that the bremsstrahlung emission also peaks at around the same frequencies as that of the two shocks, confirming that
the quasi periodicity in the emission is due to the quasi periodicity of the two shocks.

In presence of such a dynamical disk, it is intriguing to investigate the time variation of the amount of matter and angular momentum consumed by the black hole.
Let us define the mass loss parameter, or the ratio of the rate of mass cannibalized by the black hole to the rate of mass injected, as ${\dot M}/{\dot M}_{inj}$, where ${\dot M}=(\rho v_r r^2)|_{sink}$
and ${\dot M}_{inj}=(\rho v_r r^2)|_{r_{inj}}$. The angular momentum loss rate is defined as
${\dot L}/{\dot L}_{inj}$, where ${\dot L}={\dot M}l_{sink}$ and ${\dot L}_{inj}={\dot M}_{inj}
l_{inj}$. The average specific angular momentum of the disk is defined as
\begin{equation}
<l>=\frac{\int l dr}{\int dr}
\end{equation}

In Figure 9, we plot the mass-loss parameter (top panel), angular momentum loss rate (middle),
and the average angular momentum of the disk (bottom) as a function of time. The profiles of the mass-loss parameter and the angular momentum loss rate are similar to that of the bremsstrahlung emission rate.
Since the distribution of $\rho$ peaks when the shock is at the minima, the peaks of the mass-loss parameter and the angular momentum loss rate coincide with the peak of the bremsstrahlung emission.
As the shock recedes, $v_r$ and $\rho$ decreases resulting in matter getting accumulated in
the disk \ie  ${\dot M}/{\dot M}_{inj}<1$. As the shock contracts it squeezes more matter into the black hole (accumulated in the expansion phase) than it is being supplied, therefore ${\dot M}/{\dot M}_{inj}>1$.
It is to be noted that in this particular case, the disk prefers to stay in the state where ${\dot M}/{\dot M}_{inj} \lsim 1$. The angular momentum loss rate
follows ${\dot M}/{\dot M}_{inj}$. Interestingly, the maxima of the average angular momentum of the disk coincides with the minima of the emission, mass-loss rate and the angular momentum loss rate. The bottom panel of Figure 9 suggests that if the average angular momentum of the disk increases, then $v_r$
should decrease in a large region of the disk, which should reduce the rate of matter actually accreted onto the black hole. And the average angular momentum ($<l>$) of the disk increases with the increase of
the peak and the width of the angular momentum distribution of the disk, which corresponds to the dips in emission, mass-loss parameter and angular momentum loss rate.
Although $<l>$ oscillates with the same period as that of shock, but the disk interestingly prefers to stay in the state $<l>$ is greater than $l_{inj}$.
Since the disk itself is oscillating, all these flow parameters should oscillate
with the same period. And indeed the bremsstrahlung emission, the mass loss rate, the angular momentum loss rate etc all oscillate with the same period of shock oscillation.

\subsubsection{Shock oscillation for higher viscosity}

The dynamics of the disk with higher viscosity parameter is different from that due to the lower one. For higher viscosity parameter the difference in the disk dynamics will arise from more efficient angular momentum transfer as well as higher viscous dissipation of heat, even if the outer boundary condition remains the same.
In Figure 10, we have plotted the shock location with time (top) and the bremsstrahlung emission
with time (bottom) for a fluid with the same injection parameters as the inviscid flow described in \S
4.1, i.e., $r_{inj}=1000r_g$, $v_r(inj)=2.970 \times 10^{-2}c$, and $c_s(inj)=4.827 \times 10^{-3}c$
and the viscosity parameter is $\alpha=0.1$. The time variation of shock for $\alpha=0.1$ (Fig. 10)
is distinctly different from that of $\alpha=0.01$ (\ie Fig. 7). The inner shock forms, expands, and at some epoch
collides with the contracting outer shock, while at some other epoch it disappears before colliding with the outer shock. The inner shock is weaker compared to the disk with lower $\alpha$. The time evolution of the two shocks are somewhat similar to the initial phases of the shock variation for $\alpha=0.01$. Comparison of the time variation of bremsstrahlung emission with the time variation of the shock shows no correlation between shock minima and emission maxima
unlike the case for $\alpha=0.01$. In Figure 11a -- 11b, $v_r$ (dashed-dot) and $c_s$ (solid) are plotted
corresponding to the emission maxima (Fig. 11a) and emission minima (Fig. 11b). Similarly the corresponding specific angular momentum distributions are plotted for the emission maxima (Fig. 11c) and minima (Fig. 11d),
and the density too are plotted for the emission maxima (Fig. 11e) and minima (Fig. 11f). The maxima of the bremsstrahlung emission occurs when the inner shock is tending to form, while the minima occurs when the inner shock has not been formed (also refer Fig. 10). This change in the behavior of the shock
and the emission properties compared to that of the $\alpha=0.01$, actually depends on the different rate of angular momentum transfer. Since the viscosity in the present case is ten fold higher than $\alpha=0.01$, the outward angular momentum transport is very efficient. So close to the
black hole, the angular momentum rises steeply outward unlike the flow with lower viscosity (\eg Fig. 11c -- 11d may be compared with Fig. 6a -- 6d). If the shock is closer to the black hole
then the peak of the angular momentum distribution is very close to the outer shock (Fig. 11d); this causes matter to accrete more freely between the horizon and the peak of the angular momentum distribution, and hence the density is lower (Fig. 11b, 11d and 11f). This causes the emission to dip. As the shock moves out the angular momentum peak is situated further inside (Fig. 11c). And this causes the matter to madly fall inwards between the outer shock and the inner $l$ peak. As the in falling matter encounters the angular momentum pile it decelerates drastically increasing the density considerably, and hence enhances the bremsstrahlung emission (Fig. 11a, 11c and 11e). Eventually it forms an inner shock but the enhanced energy deposition in the post-inner shock region causes the inner shock to expand and thereby reducing density and emission. In this connection one may point out that the immediate post-shock (for both inner and outer shocks) region may be decelerating or accelerating (\eg Figs. 5a-5d, 11a-11b). However, it was
predicted by \citet{n92,nh94} that post-shock acceleration
and deceleration
correspond to unstable and stable shocks, respectively. The reason for this is that no
standing shock can exist in the viscous flow for the corresponding initial conditions.

In the top panel of Figure 12, we plot ${\dot M}/{\dot M}_{inj}$, and like the lower viscosity case its peak and trough coincides with that of the bremsstrahlung emission. The angular momentum loss rate
${\dot L}/{\dot L}_{inj}$ (middle) also follows the pattern of ${\dot M}/{\dot M}_{inj}$.
Since the angular momentum distribution is higher during the peak emission,
the average angular momentum of the disk $<l>$ (bottom) peak coincides with the emission peak.
Moreover, the ${\dot M}/{\dot M}_{inj}<1$ most of the time, which means because of higher angular momentum most of the matter that are being injected in to the disk, are not being consumed by the black hole, which is indicated by the fact that $<l>$ is significantly higher than $l_{inj}$.

The role of viscosity can be ascertained if one compares the viscous time with the advection time scale.
The viscous time-scale may be defined as
\begin{equation}
 \tau_{\rm vis}=\int^{r_{sh}}_{r_{ci}}\frac{r}{\nu}dr,
\end{equation}
where, $\nu=\mu/\rho$
and the advection time scale
\begin{equation}
 \tau_{\rm ad}=\int^{r_{sh}}_{r_{ci}}\frac{dr}{|v_r|}
\end{equation}
The time-variation of $\tau_{\rm vis}$ closely follows the corresponding variation of the shock location,
with minima and maxima of each, coinciding with the other at exactly the same time.
When compared for $\alpha=0.01$ case,
at the shock minima, $\tau_{\rm vis}$ and $\tau_{\rm ad}$ are comparable and we see the shock front is pushed outward. At the shock maxima $\tau_{\rm ad} << \tau_{\rm vis}$, \ie advection dominates, consequently the shock front hurls inwards contracting significantly from few $\times 100~r_g$ to few $\times 10~r_g$. Hence the interplay between these two timescales, sustains the
oscillation. The general pattern of the temporal behavior of the time scales and the relation to the shock oscillation for the flow with $\alpha=0.1$ is similar to that with the flow $\alpha=0.01$. However, $\tau_{\rm ad}$ and $\tau_{\rm vis}$ for $\alpha=0.1$ are roughly comparable and hence oscillations do not saturate.
\citet{ft00} investigated for Bondi flow that entropy-acoustic cycles may
sustain shock oscillation if $c_s(r_{ci})/c_s(r_{sh})\gg 1$.
In the simulations we have run, ${\rm max}~[c_s(r_{ci})/c_s(r_{sh})] \sim 10$ but generally the ratio is less
in most of the times. Hence the effect of entropy-acoustic cycles in regulating
shock oscillation is probably moderate in our case. A multi-dimensional simulation may give a
more definitive answer.

\section{SUMMARY AND DISCUSSION}

This paper is intended to study the time-dependent simulations of large amplitude oscillations of advective, viscous, sub-Keplerian disks, to complement earlier works of studying low amplitude oscillations undertaken by Molteni and his collaborators \citep{lmc98}. As an improvement we have employed a new code which uses the Lagrangian TVD/remap approach. This code strictly conserved the angular momentum without viscosity, and reduced the numerical dissipation considerably (\eg \S 3). Tests showed that the shock capturing capability of this code is better than both standard Eulerian code and Lagrangian SPH code (\eg Fig. 1), and followed the angular momentum transfer of the viscous, subsonic analytical solution extremely well (\eg Fig. 2).

Oscillation of accretion shock was borne out by the different rates of angular momentum transfer
across the shock and the heat dissipated due to the presence of viscosity. It has been shown that in presence of low viscosity parameter ($\alpha=0.003$), the shock front of a disk, with the same initial and boundary conditions as those of the inviscid case, did tend to expand and
settled at a larger distance from the disk (Fig. 4b). For an even higher viscosity
($\alpha \gsim 0.005$), the rate of
angular momentum transfer was higher, which caused a faster rate of shock front expansion.
As the shock front exceeded a possible equilibrium position it started to oscillate (Fig. 4c).
However, it is to be remembered that the value of the critical viscosity parameter ($\alpha \sim 0.005$ in the present case) is not sacrosanct,
but actually depends on the initial condition. For example, it has been shown that the critical viscosity
parameter will be higher for flows with lower angular momentum, while for a fluid with higher initial
energy the critical viscosity parameter will be lower \citep[see,][]{cd04}. Hence, if proper initial
condition is used then a stable shock is expected to form for higher
viscosity parameters (\ie $\alpha \sim 0.1$ -- $0.2$) too, investigation of which, however, is not the point of interest for the present paper.

A detailed study of the disk dynamics was conducted for reasonably high viscosity
(\ie $\alpha=0.01,~\&~0.1$). For $\alpha=0.01$ the shock oscillation amplitude was found to be quite high $\gsim 100r_g$. This resulted in a large sound speed gradient in the post-shock subsonic flow. In case of large amplitude shock oscillation, the rate of outward angular momentum transport in a region closer to the inner
sonic point was shown to be much higher compared to the rate of angular momentum transport near the shock.
As a result, our simulation showed that the angular momentum to be piled up in an intermediate region between the shock and the inner sonic point.
The expanding shock also increased the inflow velocity in the immediate post-shock region only to be
decelerated by the extra centrifugal pressure due to the piled-up angular momentum further inside the disk (\eg Fig. 6a -- 6d). The inflow velocity in the post shock disk may be increased to the extent that it may again become supersonic, then the resistance from the excess centrifugal pressure from the piled-up angular momentum distribution may cause the formation of inner shock. In case of moderately high $\alpha$, the distance between the peak of the angular momentum distribution and the outer shock
is large enough to allow for the $v_r$ to become supersonic again and enhanced the possibility to form the inner shock.
It is to be noted that the amplitude of shock oscillation will possibly be lesser for multi-dimensional simulation.
Viscosity is more active in the post-shock disk, and hence the extra centrifual force due to the piled angular momentum and the heat dissipated by viscosity both actively take part in shock oscillation.
However, in case of realistic accretion flow, a part of viscous heat dissipated in the post-shock disk will also be spend to puff it up, which would imply less outward push on the shock surface. Hence for a flow with same injection and viscosity parameters, the oscillation amplitude for a multi-dimensional
disk is expected to be lesser compared to a purely conical flow. Consequently, the critical viscosity above
which the disk becomes oscillatory will also be higher.

The time evolution of shocks for higher viscosity was shown to be distinctly different from that of the lower one. The inner shock was weaker and more sporadic for a disk with $\alpha=0.1$.
The main reason was because of the higher rate of angular momentum transport. Even when the shock was around $100r_g$, highly efficient angular momentum transport created a smooth increase of angular momentum, which only peaked closer to $r_{sh}$. As shock expanded $v_r$ increased, but the opportunity to become supersonic was minimized since the peak of the $l(r)$ was closer to the shock.
Hence the inner shock, if formed at all, was weaker. However, since shock amplitude for $\alpha=0.1$
was much larger than the case with $\alpha=0.01$, with time the formation of inner shock became more regular, and the behavior was more similar to that of $\alpha=0.01$.

The oscillatory motion of the shock induced oscillation in all the disk parameters like emission, rate of matter consumed by the black hole, the rate of angular momentum consumed by the black hole, and
the average angular momentum of disk. All these parameters did oscillate with the same period as that of the shock.
The disc oscillation started with $\alpha \gsim 0.005$. Considering
$M_{BH}=10~M_{\odot}$, for $\alpha=0.005$ the oscillation frequency of the outer shock was $5~Hz$, and the inner shock $10~Hz$, for $\alpha=0.006$ the frequencies were $1~Hz$ and $3~Hz$ respectively, and for $\alpha=0.01$ the two frequencies were $0.125~Hz$ and $0.25~Hz$. Hence one may conclude that
apart from the dependence of the oscillation frequency on injection parameters, the QPO frequency definitely decreases with increasing viscosity and vice-versa. Observationally, GRO J1655-40
exhibits a rise in QPO frequency in its rising state and a fall in QPO frequency in its declining
phase in 2005 \citep{cdnp08}.
\citet{cdp09} plotted the
QPO frequency for the object XTE J1550-564 in 1998 burst phase. They showed that in the rising phase
of the outburst, the low frequency QPO increases from $0.08 Hz$ to $13.1~Hz$ and then
starts to decrease in the declining phase before disappearing. Such rise and fall of QPO frequencies may be explained by the change in shock oscillation frequency due to the
change of the net viscosity of the disk.

In presence of viscosity a positive angular momentum gradient \ie $dl/dr \geq 0$ helps in outward transport of angular momentum. However, a negative gradient may trigger inward transport of angular momentum. The $dl/dr<0$ condition was attained in the disk in at least two locations, at the outer shock front and just behind the peak of the specific angular momentum distribution. Those regions were subject to the rotational instability. $dl/dr<0$ caused the average angular momentum $<l>$ of the disk to increase, and hence the period and the amplitude of the shock oscillation to increase too.
This is less perceptible for lower $\alpha$ and the shock oscillation achieved
quasi-saturation, but for $\alpha=0.1$ the shock went outside the computation domain.
We repeated the simulation with $\alpha=0.3$ (not presented in the paper) and in this case too the shock went outside the domain,
although formation of inner sonic point and oscillation of the two shocks were observed too.

In case of multi-dimensional simulations, a part of the post shock matter would have ejected along the vertical
direction in the form of winds, which would have carried away a part of the angular momentum, such that the increase of $<l>$ may have been arrested for higher $\alpha$.
This would have meant that the shock oscillation may saturate for $\alpha \geq 0.1$. Hence, we conjecture that the non-saturation of shock oscillation for $\alpha\gsim 0.1$, could be an artifact of one-dimensional
simulation. We will test it in a future work using multi-dimensional simulations.

\acknowledgments{SL was supported in part by Basic Science Research Program through the National Research Foundation of Korea (NRF) funded by the Ministry of Education, Science and Technology (2010-0004738).
DR was supported in part by National Research Foundation of Korea through grant KRF-2007-341-C00020.

\begin{figure}
\plotone{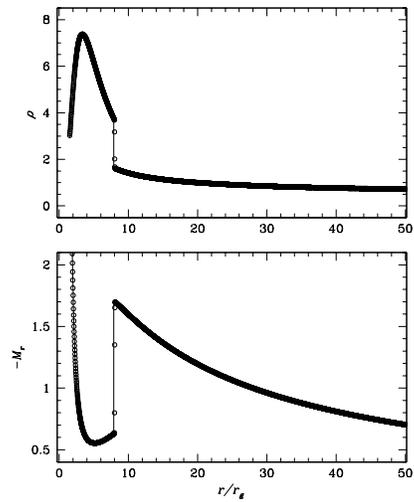} \caption{Comparison of numerical solution with
analytic solution for a one-dimensional accretion flow in
cylindrical geometry that allows a standing shock. The plots show
the density ($\rho$) and the radial Mach number ($M_r$). The solid
curves represent the analytical solution and the open circles
represent the numerical solution with 2048 uniform grid cells.}
\end{figure}

\begin{figure}
\plotone{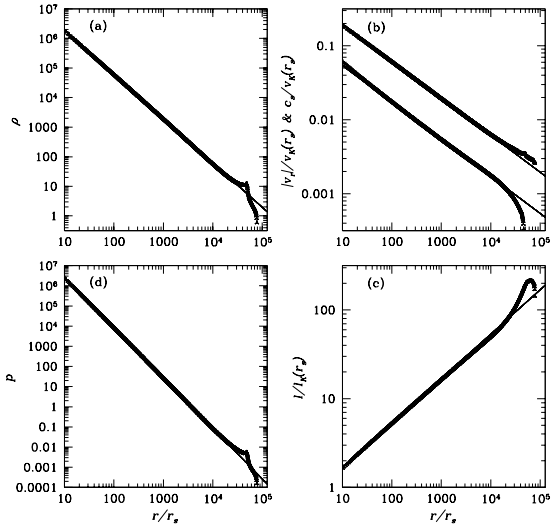} \caption{Test of ADAF with $\gamma$ = 4/3
and $\alpha$ = 0.3 under the Newtonian potential. The solid
lines represent the analytical self-similar solution, while the dots
represent the numerical solution. The density $\rho$ (a), and radial velocity $v_r$ (lower curve of (b))
\& adiabatic sound speed $c_s$ (upper curve of (b)), specific angular momentum $l$ (c), pressure
$p$ (d), are shown clockwise.}
\end{figure}

\begin{figure}
\plotone{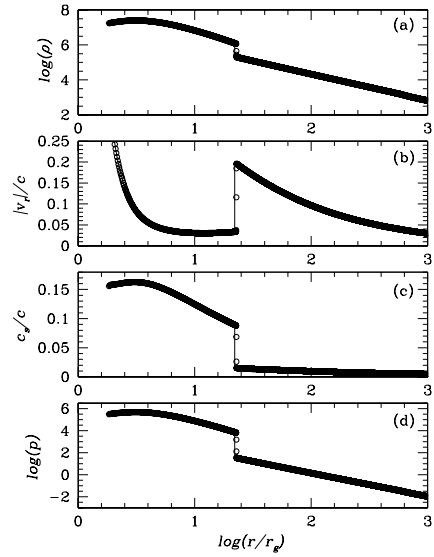} \caption{Comparison the analytical (solid) with the
numerical solution (open circles) of hydrodynamical accretion shock for transonic flow
with $\gamma = 1.4$ and $l = 1.8 c r_g$. The injection radius $r_{inj}=1000r_g$,
and at $r_{inj}$ the radial velocity
$v_r(inj)=2.970 \times 10^{-2}c$ and sound speed $c_s(inj)=4.827 \times 10^{-3}c$.}
\end{figure}

\begin{figure}
\plotone{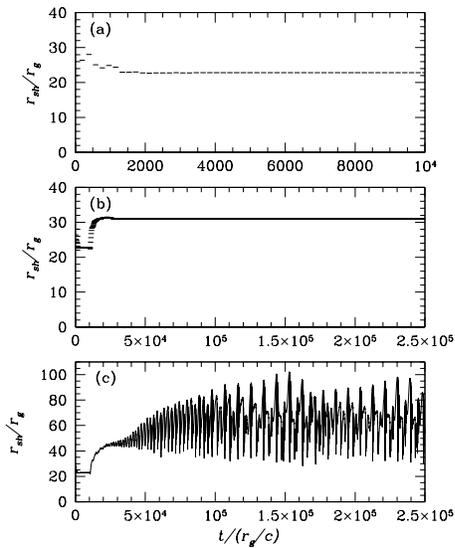} \caption{Comparison of the shock location $r_{sh}$ for
$\alpha=0.0$ (a), $\alpha=0.003$ (b) and $\alpha=0.006$ (c). The injection radius is $r_{inj}=1000r_g$,
the injected radial velocity
$v_r(inj)=2.970 \times 10^{-2}c$, sound speed $c_s(inj)=4.827 \times 10^{-3}c$, and angular momentum
$l_{inj} = 1.8 c r_g$. The adiabatic index is $\gamma = 1.4$.}
\end{figure}

\begin{figure}
\plotone{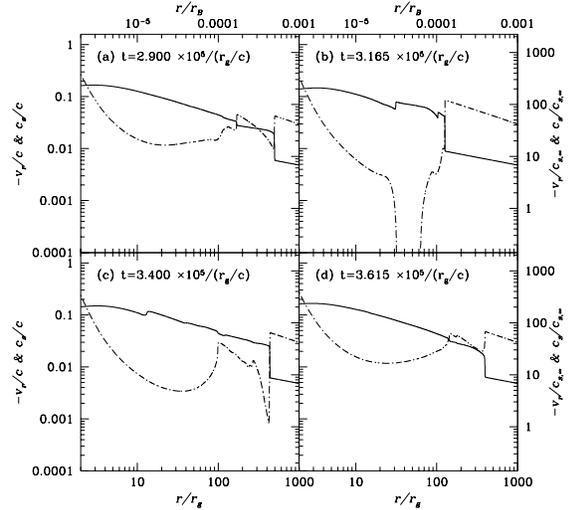} \caption{Four snapshots of radial velocity (dashed-dot) and sound speed (solid) of a viscous
fluid of $\alpha=0.01$ and $\gamma=1.4$. The injection parameters are $r_{inj}=1000r_g$,
$l_{inj}=1.8cr_g$,
$v_r(inj)=2.970 \times 10^{-2}c$, and $c_s(inj)=4.827 \times 10^{-3}c$.
The time sequence goes as (a) $\rightarrow$ (b)
$\rightarrow$ (c) $\rightarrow$ (d).
}
\end{figure}

\begin{figure}
\plotone{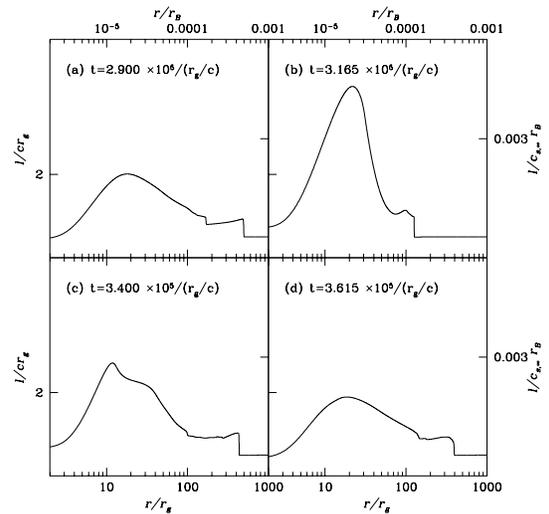} \caption{Specific angular momentum at the four
different snapshots in the same simulation as in Fig. 5. The time
sequence goes as (a) $\rightarrow$ (b) $\rightarrow$ (c)
$\rightarrow$ (d).
}
\end{figure}

\begin{figure}
\plotone{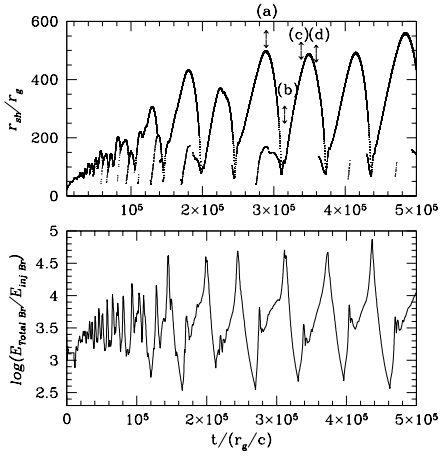} \caption{Shock position (top) and bremsstrahlung emission (bottom) as a function of time in the same simulation as in Fig. 5. The upper curve in the top panel is the outer shock and the lower curve is the inner shock. The snap shots of time in Fig. 5a -- 5d are marked on the top panel
as a -- d.}
\end{figure}

\begin{figure}
%
\plotone{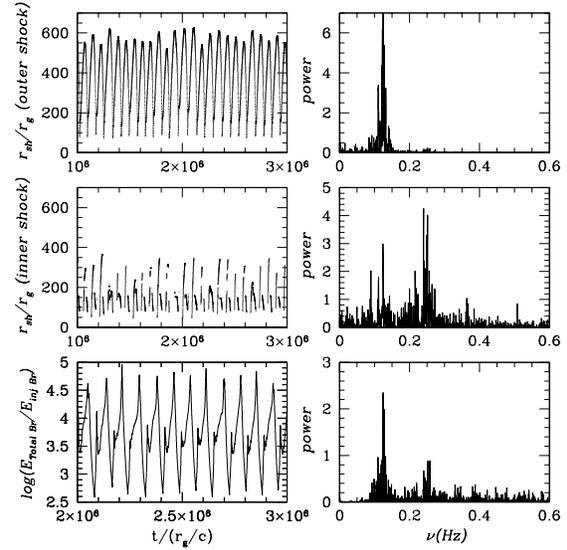} \caption{Left panels: Outer shock (top), inner shock (middle) and the
bremsstrahlung emission (bottom) as a function on of time. Right panels: The power density spectra
of the outer shock (top), inner shock (middle) and the bremsstrahlung emission (bottom). Same simulation as in Fig. 5.}
\end{figure}

\begin{figure}
\plotone{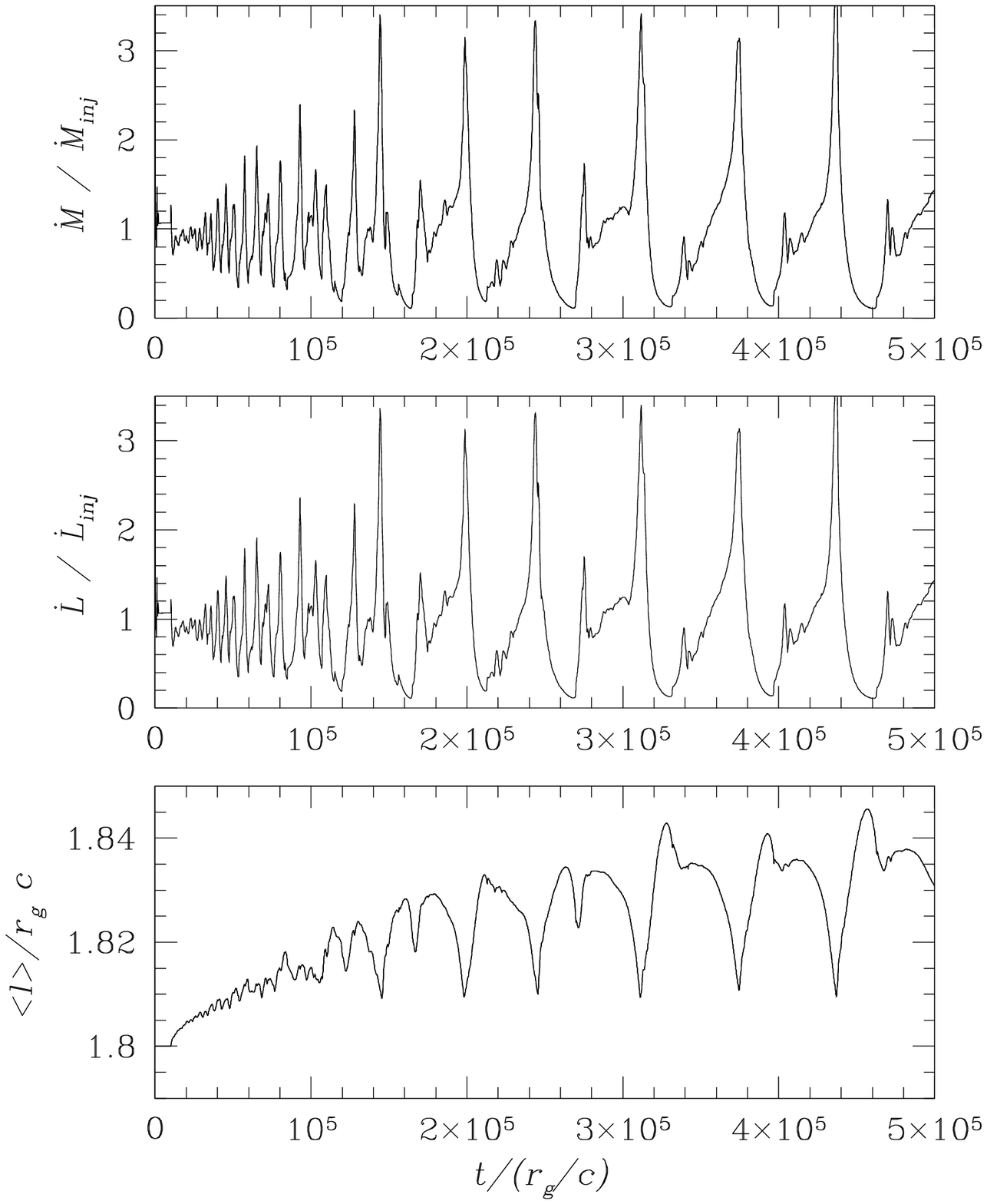} \caption{Mass accretion rate (top), angular momentum loss rate (middle) and averaged
specific angular momentum (bottom) in the post shock region of the outer
shock as a function of time in the same simulation as in Fig. 5.}
\end{figure}

\begin{figure}
\plotone{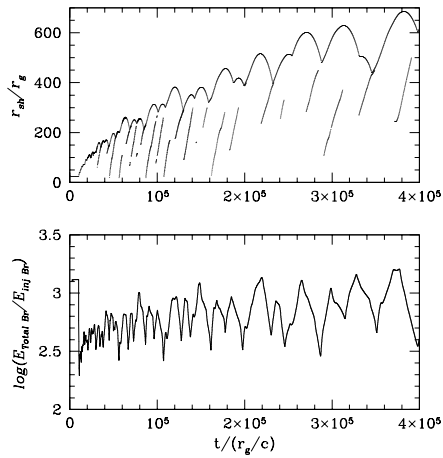} \caption{Shock position (top) and bremsstrahlung emission (bottom) as a function of time. The upper curve in the top panel is the outer shock and the lower curve
is the inner shock. The simulation is for $\alpha=0.1$ and $\gamma=1.4$, and injection parameters
$r_{inj}=1000r_g$, $l_{inj}=1.8cr_g$,
$v_r(inj)=2.970 \times 10^{-2}c$, and $c_s(inj)=4.827 \times 10^{-3}c$.}
\end{figure}

\begin{figure}
\plotone{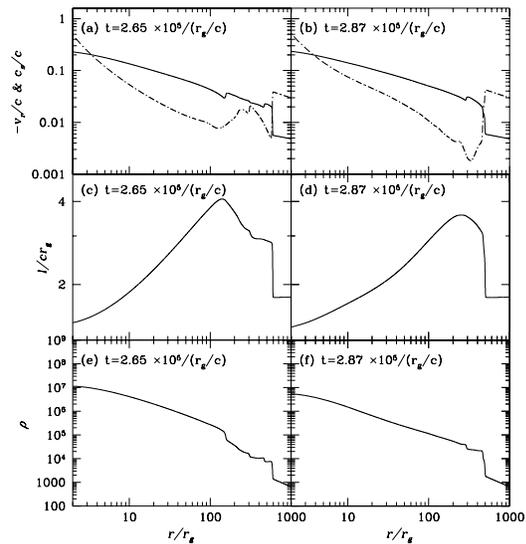} \caption{Radial velocity (dashed-dot) and sound speed (solid) of viscous
fluid for time $t=2.65{\times}10^5\tau_g$ (a) and $t=2.87{\times}10^5\tau_g$ (b).
Specific angular momentum is plotted for time $t=2.65{\times}10^5\tau_g$ (c) and $t=2.87{\times}10^5\tau_g$ (d). Density in arbitrary units is plotted  for $t=2.65{\times}10^5\tau_g$ (e) and $t=2.87{\times}10^5\tau_g$ (f).
The simulation is for $\alpha=0.1$, and $\gamma=1.4$. The injection parameters are $r_{inj}=1000r_g$, $l_{inj}=1.8cr_g$,
$v_r(inj)=2.970 \times 10^{-2}c$, and $c_s(inj)=4.827 \times 10^{-3}c$.}
\end{figure}

\begin{figure}
\plotone{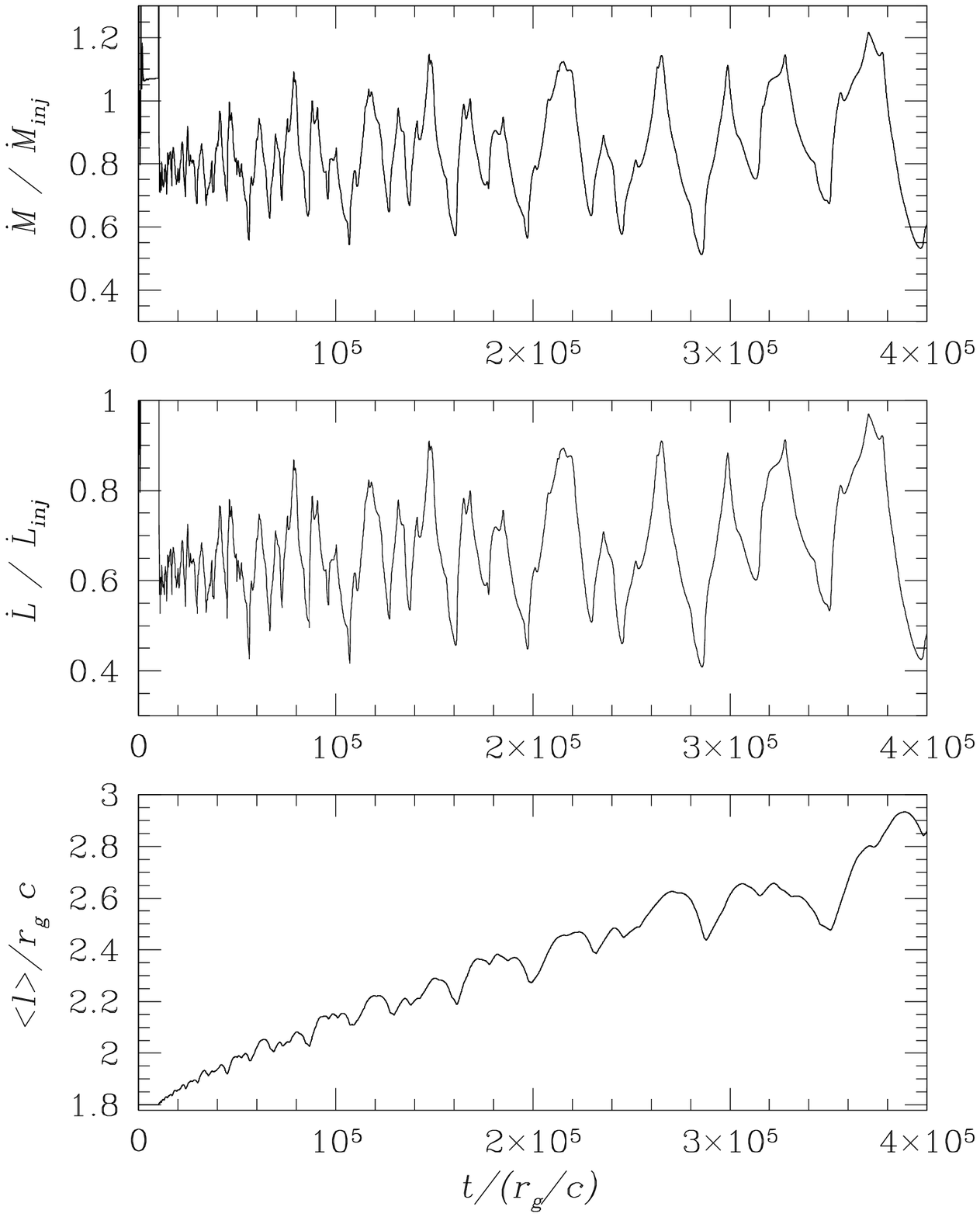} \caption{Mass accretion rate (top), angular momentum loss rate (middle) and averaged
specific angular momentum (bottom) in the post shock region of the outer
shock as a function of time in the same simulation as in Fig. 10.}
\end{figure}

\end{document}